\documentstyle[12pt]{article}
\begin{document}
\begin{titlepage}

\vspace{1cm}

\begin{center}
{\Large \bf  Budker Institute of Nuclear Physics}
\end{center}

\vspace{0.5cm}

\begin{flushright}
BINP 97-97\\
December 1997
\end{flushright}

\vspace{1cm}

\begin{center}
{\Large \bf Tail of Gravitational Radiation\\
and Coulomb Final-State Interaction }
\end{center}

\begin{center}
I.B. Khriplovich\footnote{khriplovich@inp.nsk.su} and
A.A. Pomeransky\footnote{pomeransky@inp.nsk.su}
\end{center}
\begin{center}
Budker Institute of Nuclear Physics\\
630090 Novosibirsk, Russia
\end{center}

\bigskip

\abstract{We present a simple intuitive derivation of the corrections
to the intensity of gravitational radiation due to the so-called tail effect.}

\vspace{3.5cm}

\end{titlepage}

\begin{figure}[b]
\begin{picture}(230,125)
\put(0,0) {\begin{picture}(230,100)
            \thicklines
            \put(170,80){\line(1,0){25}}
            \put(80,80){\line(1,0){90}}
            \put(50,80){\line(1,0){30}}
            \put(100,80){\line(1,1){40}}
            \put(130,110){\line(1,1){15}}
            \put(100,80){\line(1,-1){40}}
            \put(130,50){\line(1,-1){15}}
            \multiput(140,80)(0,-10){4}{\line(0,-1){5}}
            \put(120,107){$\tilde{\nu}$}
            \put(160,85){$p$}
            \put(120,48){$e$}
            \put(80,85){$n$}
            \put(20,0){Fig. 1. Coulomb correction to $\beta$-decay.}
            \end{picture}}

\put(230,0) {\begin{picture}(230,100)
            \put(160,25){\line(0,1){5}}
            \put(95,100){\oval(10,20)[bl]}
            \put(95,80){\oval(10,20)[tr]}
            \put(110,75){\oval(20,20)[bl]}
            \put(110,60){\oval(20,10)[tr]}
            \put(130,55){\oval(20,20)[bl]}
            \put(130,40){\oval(20,10)[tr]}
            \put(150,35){\oval(20,10)[bl]}
            \put(150,25){\oval(20,10)[tr]}
            \multiput(140,35)(-20,20){3}{\line(0,1){5}}

            \thicklines
            \put(170,100){\line(1,0){25}}
            \put(50,100){\line(1,0){120}}
            \multiput(140,40)(0,10){6}{\line(0,1){5}}
            \put(20,0){Fig. 2. Newton correction to graviton emission.}
            \end{picture}}

\end{picture}

\end{figure}

A special correction to the radiation intensity due to the so-called tail of
radiation in gravitating systems,
was pointed out long ago \cite{bd}. The effect was
dicussed afterwards in many articles, all corresponding references being too
numerous to
be presented in this short note. In particular, it was recognized that the
phenomenon is due in fact to the gravitational attraction in the final state
between the emitted radiation and emitting system. Still, in our opinion a
simple picture of the effect is lacking up to now. Especially mysterious looks
the dependence of the correction on $v/c$. As distinct from usual relativistic
corrections, which
contain only even powers of $c^{-1}$, this one starts with $c^{-3}$.

\begin{figure}[b]
\begin{picture}(230,100)
\put(0,0) {\begin{picture}(80,100)
            \multiput(42,73)(10,0){3}{\oval(5,5)[b]}
            \multiput(47,73)(10,0){3}{\oval(5,5)[t]}
            \thicklines
            \multiput(0,50)(0,50){2}{\line(1,0){30}}
            \multiput(50,50)(0,50){2}{\line(1,0){30}}
            \multiput(30,50)(0,50){2}{\line(1,0){20}}
            \multiput(40,50)(0,10){5}{\line(0,1){5}}
            \put(0,15){Fig. 3. Diagram with graviton emission}
            \put(0,0){from two $g_{00}$, Newton, lines.}
            \end{picture}}
\put(230,0) {\begin{picture}(80,100)
            \multiput(43,72)(10,0){3}{\oval(5,5)[b]}
            \multiput(48,72)(10,0){3}{\oval(5,5)[t]}
            \thicklines
            \multiput(0,50)(0,50){2}{\line(1,0){30}}
            \multiput(50,50)(0,50){2}{\line(1,0){30}}
            \multiput(30,50)(0,50){2}{\line(1,0){20}}
            \multiput(40,50)(0,10){2}{\line(0,1){5}}
            \multiput(39,70)(0,5){6}{.}
            \put(0,15){Fig. 4. Diagram with graviton emission}
            \put(0,0){from $g_{00}$ and $g_{0n}$ (dotted line) lines.}

        \end{picture}}

\end{picture}
\end{figure}

We present here an explanation of the effect based on its close analogy
with the Coulomb
final-state interaction. It is well-known (see for instance the textbook
\cite{ll}) that the Coulomb final-state attraction between the particles
modifies the probability by a factor
\begin{equation}\label{psi0^2}
{{\xi} \over {1-\exp(-\xi)}}\,,
\end{equation}
where
\begin{equation}\label{xi}
\xi={{2 \pi Z_1 Z_2 e^2} \over {\hbar v}},
\end{equation}
and $Z_1 e$ and $Z_2 (-e)$ are the charges of the respective particles, $v$
is their relative velocity. As an example one can consider the neutron
$\beta$-decay with the Coulomb interaction between the electron and proton in
final state (see Fig. 1, where the dashed line refers to the multiple
Coulomb exchange).

Let us consider now the process we are interested in, that of the graviton
emission. It can be presented by a similar diagram, Fig. 2, where the dashed
line refers to the final-state gravitational attraction between the emitted
graviton (wavy line) and the system itself. There is an obvious analogy with
the previous process. The correspondence is as follows:
\[ v \rightarrow c,\;\;\;  Z_1 Z_2 e^2 \rightarrow 2 k M \mu. \]
Here $k$ is the Newton gravitational constant, $M$ is the mass of the radiating
system, $\mu = \hbar \omega/c^2$ is the gravitating mass of emitted radiation,
$\omega$
being its frequency. The overall factor 2 accounts for the well-known fact
that the gravitational attraction for a massless particle is twice as large
as that for a massive nonrelativistic one. So, in the gravitational problem
we are interested in, the tail effect is described by the same formula
(\ref{psi0^2}) where now
\begin{equation}\label{xi1}
\xi={{4 \pi k M \omega} \over c^3}.
\end{equation}

We wish to emphasize that both corrections due to the final-state
interaction, the usual Coulomb correction and the gravitational "Newton"
one, we are interested in here, are of a universal nature in the following
sense. Both enter as an overall factor at the probability or intensity
calculated without them.

To first order in $\xi$ the relative final-state correction to
the radiation intensity is
\begin{equation}\label{xi2}
{{2 \pi k M \omega} \over c^3},
\end{equation}
in agreement with the original result [3--5].
For the dipole electromagnetic radiation its frequency $\omega$ coincides
with the revolution frequency $\omega_0$. For the quadrupole gravitational
radiation $\omega = 2 \omega_0$.

Obviously, one can obtain corrections of higher order in $\xi$, just
expanding formula (\ref{psi0^2}). In particular, we reproduce easily
the next-order correction derived originally in \cite{TSTS}:
\begin{equation}
\frac{1}{3}\left({{4 \pi k M} \omega_0 \over c^3}\right)^2.
\end{equation}

Since the wavelength of radiation by a nonrelativistic system is large as
compared to the size of the system, in a two-body problem $M$ is nothing but
the sum of the masses of components:
\[ M = m_1+ m_2. \]
Thus, parameter $\xi$ in a two-body problem becomes
\begin{equation}
{{4 \pi k (m_1+ m_2)} \omega_0 \over c^3}.
\end{equation}

One may wonder about the contributions to the intensity of radiation due to
diagrams different from Fig. 2. The
question is conveniently discussed in the case of a two-body problem. The
diagram presented in Fig. 3, where the graviton interacts with two
$g_{00}$ lines, Newton ones, is in fact taken care of in the standard
transformation to the quadrupole formula when starting from the spatial
components of the stress-energy tensor $T_{mn}$ ($m,n=1,2,3$) which are
the true source of the emitted graviton.
Analogously, diagram of the type Fig. 4, with dotted line corresponding to the
propagation of $g_{0n}$ field, is accounted for by usual relativistic
corrections to the simple quadrupole formula. All other diagrams either cancel
or generate corrections of higher order in $1/c$.

\bigskip
\bigskip

We are grateful to R.A. Sen'kov for useful discussions and to B.S. DeWitt for
the advice to publish this note.

\end{document}